\documentclass[authoryear,round,12pt]{article}
\usepackage{color}
\usepackage{pslatex}
\usepackage{graphicx}
\usepackage{setspace}
\usepackage{natbib}
\usepackage{amsmath}
\usepackage{subfigure}
\newcommand{\myabstract}{We present results of a forecast initiated
  Week 49 (beginning December 9, 2012) of the 2012-2013 influenza
  season for municipalities in the United States.  The forecast was
  made on December 14, 2012.  Results from forecasts initiated the two
  previous weeks (Weeks 47 and 48) are also presented.  Also results
  from the forecast generated with the SIRS model without absolute
  humidity forcing (no AH) are shown.}

\begin{document}

%
\title{\textbf{\large{Week 49 Influenza Forecast for the 2012-2013
      U.S. Season}}
%
%
\author{\textsc{Jeffrey Shaman}
                                \thanks{\textit{Corresponding author address:} 
                                Jeffrey Shaman, Department of
                                Environmental Health Sciences, Mailman
                                School of Public Health, Columbia
                                University, 722 West 168th Street,
                                Rosenfield Building, Room 1104C, New
                                York, NY 10032. 
                                \newline{E-mail:
                                  jls106@columbia.edu}}\quad\textsc{}\\
\centerline{\textit{\footnotesize{Department of Environmental Health Sciences,
    Mailman School of Public Health, Columbia University, New York, New York}}}
\and
\centerline{\textsc{Alicia Karspeck}} \\
\centerline{\textit{\footnotesize{Climate and Global Dynamics
      Division, National Center for Atmospheric Research, Boulder, Colorado}}}
\and 
\centerline{\textsc{Marc Lipsitch}} \\
\centerline{\textit{\footnotesize{Center for Communicable Disease
      Dynamics, Harvard School of Public Health, Harvard University,
      Boston, Massachussetts}}}
}}

%



\maketitle

{
\begin{abstract}
\myabstract
\end{abstract}
}


\section{Retrospective Forecast}
\label{sec:retrofore}

Retrospective forecast skill is calculated for individual cities, as
well as for census region in aggregate, and all cities (excluding the
pandemic years 2008-2009 and 2009-2010, which will need to be handled
separately in the future).  The forecast methods are similar to those
described in \cite{Shaman-Karspeck-2012:forecasting}.  Based on the
relationship between prediction accuracy and ensemble spread of these
retrospective forecasts we can assign calibrated confidences to our
current predictions.

Some cities work well in isolation (St. Louis, NYC--not shown), others
do not.  The question is whether the good and bad cities should simply
be aggregated by region, which would suggest that the
predictability is really the same among them, but the
sample size for an individual city is too small (too few years). Or is
it that the statistics are robust and that predictability varies among
cities due to differences in local population size, population age
structure, geography and connectivity among individuals, etc.  We
don't know the answer to this question yet, so for now will give
certainty estimates based on the local/municipal record, the regional
aggregate and the national aggregate.

\begin{figure}[tbh]
\noindent\includegraphics[width=20pc,angle=0]{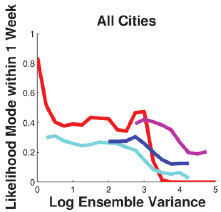}
\noindent\includegraphics[width=20pc,angle=0]{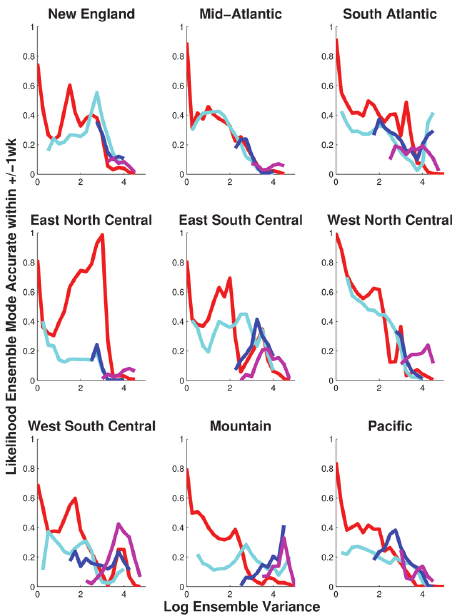}\\
\caption{Plot of ensemble mode forecast accuracy versus ensemble
  spread measured as log ensemble variance +1.  Left) 114 cities in
  aggregate.  The runs are binned in
  increments of 0.25 units and stratified by forecast lead time: 1-3
  weeks (red), 4-6 weeks (cyan), 7-9 weeks (blue), 10+ weeks
  (magenta).  Right) Same as left, but the 114 cities grouped by census region.}
\label{fig:allcities_mode_allclim}
\end{figure}

Figure \ref{fig:allcities_mode_allclim} shows the results for all
cities in aggregate using climatological AH and a factor of 5 mapping.
Overall the relationship is informative; however, for all lead times
there is a basic plateau of skill once the ensemble log variance drops
below 2.5 to 3 weeks$^2$.

When the cities are grouped by region, there is some heterogeneity.
Some regions (e.g. West North Central) show marked improvement of
forecast accuracy/skill with decreasing spread across all lead times.
Other regions show much more limited skill--the Mountain region only
has skill at 1-3 weeks, and the East North Central has problems at 1-3
weeks.  
 
\section{2012-2013 Forecast}
\label{sec:actualfore}

\subsection{Week 49 Forecast}
\label{subsec:actual49}
 
\begin{figure}[tbh]
\noindent\includegraphics[width=25pc,angle=0]{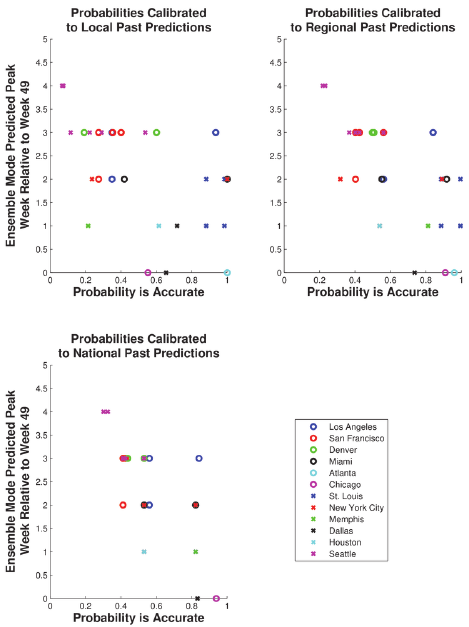}\\
\caption{Ensemble mode peak week predictions relative to Week 49 for
  12 cities plotted as a function of probability/confidence calibrated
  from historical city, regional and national prediction accuracy.
  Forecasts were initiated the beginning of Week 50 (December 9, 2012).}
\label{fig:select_wk49fore_cal}
\end{figure}

The Week 49 forecast (initiated December 9, 2012) basically stays on
track with predictions made in prior weeks.  Atlanta and Chicago are
all predicted to be at peak ($\pm1$ week) during week 49 (Figure
\ref{fig:select_wk49fore_cal}), which is the week ending December 8,
2012.  Basically, these forecasts predict no future week higher then
that latest observed week.  Dallas is forecast to predict in 0-1
weeks.  Houston and Memphis are forecast to peak in one week (the week
ending December 15, 2012).  The calibrated confidence in these
predictions is fairly high ($>50\%$, except Memphis at the municipal scale).

\begin{figure}[tbh]
\noindent\includegraphics[width=18pc,angle=0]{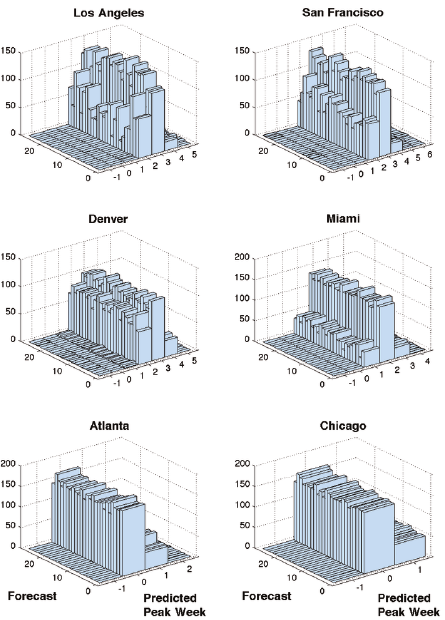}
\noindent\includegraphics[width=18pc,angle=0]{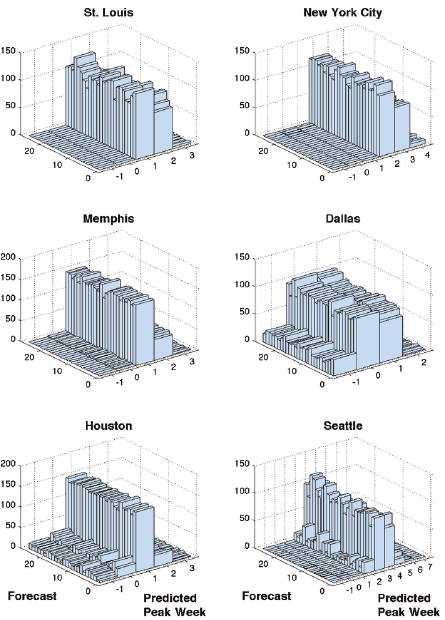}\\
\caption{Left) Histograms of the best ensemble start date trainings
  for forecasts made beginning the start of Week 50 (December 9,
  2012) for select cities.  The distributions show the ensemble spread
  among peak predictions.  Week 0 is Week 49.}
\label{fig:select_wk49fore}
\end{figure}

St. Louis is predicted to peak in 1-2 weeks.  Miami is now predicted
to peak in 2 weeks, as is New York City.  The New York City prediction
is a change of 2-3 weeks from the prior week (Week 48) prediction of
peak in 4-5 weeks.  Both Los Angeles and San Francisco are forecast to
peak in 2-3 weeks.  Denver is predicted to peak in 3 weeks (with about
$50\%$ confidence).  Seattle is predicted to peak in 3-4 weeks.
Figure \ref{fig:select_wk49fore} shows histograms of these predictions.

\begin{figure}[t]
\noindent\includegraphics[width=18pc,angle=0]{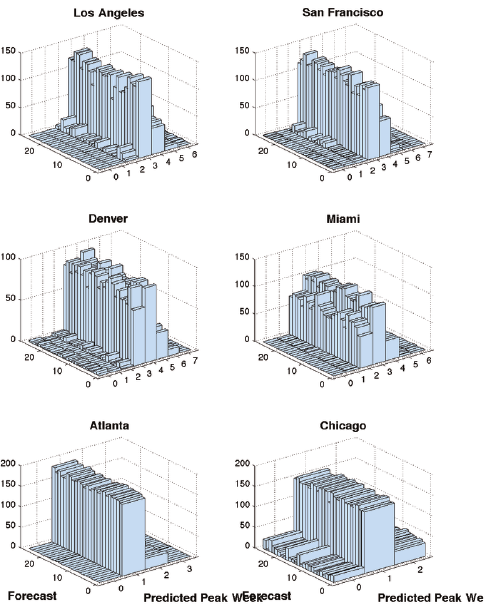}
\noindent\includegraphics[width=18pc,angle=0]{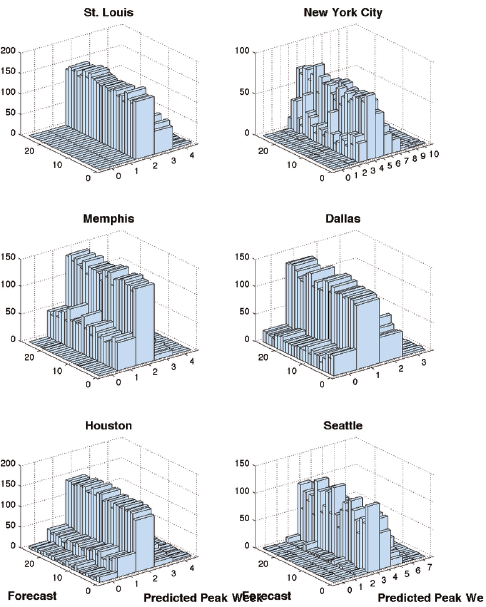}\\
\caption{Left) Histograms of the best ensemble start date trainings
  for forecasts made beginning the start of Week 49 (December 2,
  2012) for select cities.  The distributions show the ensemble spread
  among peak predictions.  Week 0 is Week 48.}
\label{fig:select_wk48fore}
\end{figure}

Most of these changes are shifts in the prediction of 1 week from the
prior week prediction, indicative of a similar tracking of outbreak
evolutions.  New York City had a larger shift of 2-3 weeks.  (Indeed,
two weeks ago, New York City had a peak predicted to be 6-7 weeks in
the future; so the forecast trajectory has shifted.)  From the
histograms, it can be seen that overall for New York City, there is
less spread among the forecasts within an ensemble--the histogram is
tightly spread between 2-3 week leads, whereas for the week prior
(Figure \ref{fig:select_wk48fore}) there was greater spread in the
predicted peak weeks.

\subsection{Week 47 Forecast}
\label{subsec:actual47}

The week 47 forecast (started November 25, 2012) predicts an outbreak
peak in 6-7 weeks for NYC (best at 32 week start), 3-4 weeks in Miami
(all have variance of 0.4-0.7, so 40-50\% confidence), Chicago in
3weeks (0.3, so 40-50\%), Denver in 3 weeks (0.4, 50\% confidence), LA
5 weeks (1.3-2 variance, so 20-30\% confidence based on region and
national), SF 5 weeks (1.8-2 variance, so 20-25\%), Dallas, 2 weeks
(0.25 variance 50-80\% likelihood), Houston (2-3 weeks, 0.3 variance,
50\% likelihood), Atlanta in 4 weeks (0.5 variance, 40-60\%), Memphis
in 3 weeks (0.5 variance, 40-50\%).

\begin{figure}[t]
\noindent\includegraphics[width=18pc,angle=0]{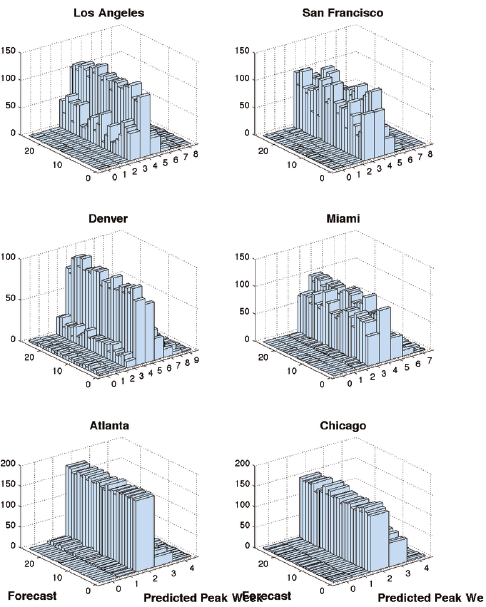}
\noindent\includegraphics[width=18pc,angle=0]{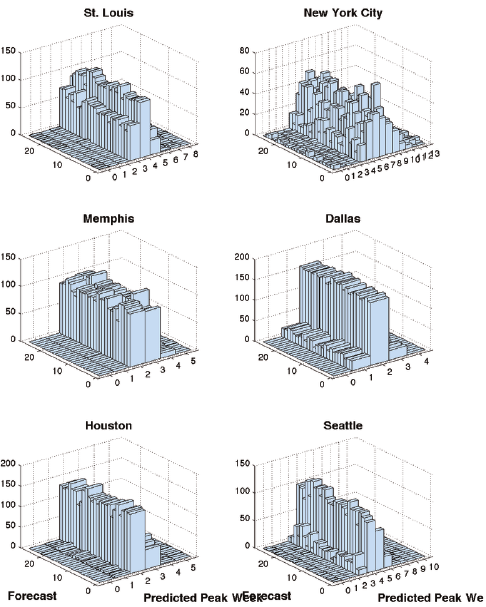}\\
\caption{Left) Histograms of the best ensemble start date trainings
  for forecasts made beginning the start of Week 48 (November 25,
  2012) for select cities.  The distributions show the ensemble spread
  among peak predictions.  Week 0 is Week 47.}
\label{fig:select_wk47fore}
\end{figure}

The distributions for these are shown in Figure
\ref{fig:select_wk47fore}.  The forecasts are, as expected that flu
will peak in the south earlier and later in the Northeast and west.

\begin{figure}[t]
\noindent\includegraphics[width=25pc,angle=0]{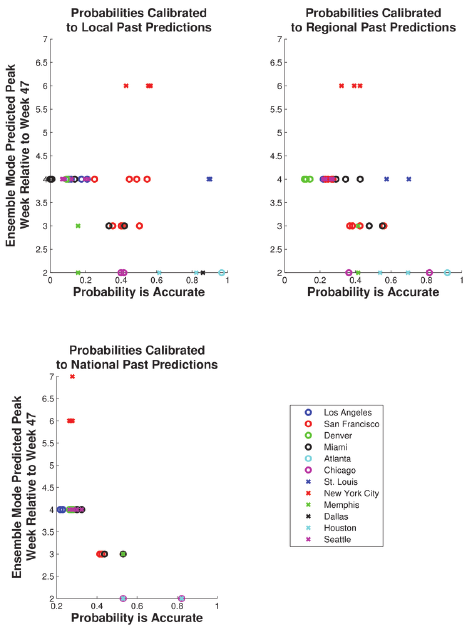}\\
\caption{Ensemble mode peak week predictions relative to Week 47 for
  12 cities plotted as a function of probability/confidence calibrated
  from historical city, regional and national prediction accuracy.
  Forecasts were initiated the beginning of Week 48 (November 25, 2012).}
\label{fig:select_wk47fore_cal}
\end{figure}

The probabilities/certainties associated with those predictions for
those cities are shown calibrated to historical city, regional and
national prediction accuracy (Figure
\ref{fig:select_wk47fore_cal}--city calibration not shown).
The Atlanta forecasts are very confident based on local and regional
calibration.  The St. Louis forecast is also very strong (and 4 weeks
ahead) based on local and regional calibration.

The 3-week lead forecast for Miami is about 40\% regardless of local,
regional or national calibration.  NYC is at 40-50\% confident
forecast based on local and regional calibration with a 6 week lead.

\subsection{Week 48 Forecast}
\label{subsec:actual48}

These forecasts are begun after assimilation of the Week 48 ILI+
observations (week ending December 1, 2012).  

\begin{figure}[t]
\noindent\includegraphics[width=25pc,angle=0]{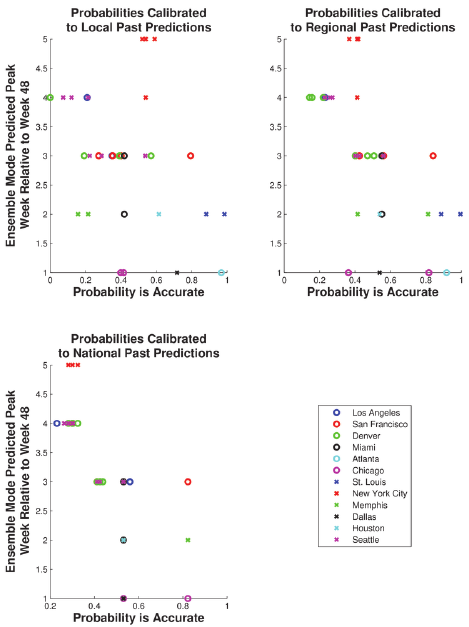}\\
\caption{Ensemble mode peak week predictions relative to Week 48 for
  12 cities plotted as a function of probability/confidence calibrated
  from historical city, regional and national prediction accuracy.
  The forecasts were initiated the beginning of Week 49 (December 2, 2012).}
\label{fig:select_wk48fore_cal}
\end{figure}

The forecasts seems to be consistent with the prior week's in that
most have shifted to be one week nearer peak (Figure
\ref{fig:select_wk48fore_cal})).  Some have jumped more,
i.e. St. Louis, which went from 4 to 2 weeks in the future, with a big
jump in certainty (~90\% probability) as calibrated from local and
regional baseline historics.  Note that all forecasts assume an error
of $\pm1$ week, so a 2-week jump may still represent accurate tracking

Basically, we see predictions of peaking in one week (the week ending
December 8, which is today) for Dallas, Chicago and Atlanta, two weeks
for St. Louis, Houston, Miami and Memphis, 3-4 weeks for San
Francisco, Los Angeles, Seattle and Denver, and 4-5 weeks for NYC--all
with varying levels of confidence, though tracking consistently with
last weeks forecast (Figure \ref{fig:select_wk47fore}).

\subsection{Week 49 Forecast -- No AH}
\label{subsec:actual49noAH}

\begin{figure}[tbh]
\noindent\includegraphics[width=25pc,angle=0]{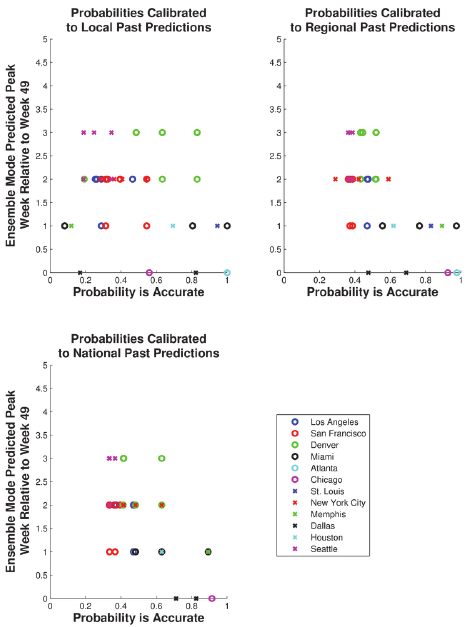}\\
\caption{Ensemble mode peak week predictions relative to Week 49 for
  12 cities plotted as a function of probability/confidence calibrated
  from historical city, regional and national prediction accuracy.
  These are derived from training and forecasts with the SIRS model
  without AH forcing. Forecasts were initiated the beginning of Week
  50 (December 9, 2012).}
\label{fig:select_wk49fore_cal_noAH}
\end{figure}

Week 49 forecasts (again initiated December 9, 2012 at the beginning
of Week 50) produce a similar range of predictions (Figure
\ref{fig:select_wk49fore_cal_noAH}).  Specifically, Dallas, Atlanta
and Chicago are predicted to peak Week 49 (zero weeks in the future).
The Dallas forecasts are not always skillful, depending on ensemble
spread and whether calibration is to local, regional or national
historical relationships between accuracy and spread; however, the
Atlanta prediction is very certain.  We'll see.  These leads agree
with the forecast made using the climatological AH forced SIRS model
(Figure \ref{fig:select_wk49fore_cal})--though Dallas has a 0-1 week
prediction in that case.

Memphis, Houston, St. Louis, and Miami are all predicted to peak in 1
week (Week 50, which is near conclusion at time of writing).  Memphis
and Houston are in agreement with the climatological AH forecasts;
St. Louis is a bit earlier having changed to 1 week from 1-2 weeks
with this form; Miami is 1 week earlier with the no AH form.  Both
shifts are within the margin of error for the predictions ($\pm1$
week).  

San Francisco and Los Angeles are predicted to peak in 1-2 weeks with
certainties between 30 and 50\%.  These are a week earlier than the
predictions with the climatological AH SIRS form (2-3 weeks).  The
latter form, produced a few higher confidence predictions for LA at 3
weeks (60-90\%, Figure \ref{fig:select_wk49fore_cal}).

New York City is predicted to peak in 2 weeks here, in agreement with
the climatological AH SIRS forecasts.  Seattle and Denver both show
2-3 weeks forecasts for peak week.  The Denver forecasts are more
certain when calibrated locally.  These forecasts are a little earlier
than the climatological AH made counterparts (by 1 and 0-1 weeks,
respectively, still in the margin of error.)

{}
{\clearpage}

\bibliographystyle{apalike}
\bibliography{week48bib}

\end{document}